# High-Rate Continuous Variable Quantum Key Distribution Based on Probabilistically Shaped 64 and 256-QAM

François Roumestan[(1)], Amirhossein Ghazisaeidi[(1)], Jérémie Renaudier[(1)], Luis Trigo Vidarte[(2)], Eleni Diamanti[(2)], and Philippe Grangier[(3)]

[(1)] Nokia Bell Labs, Paris-Saclay, route de Villejust, F-91620 Nozay, France, francois.roumestan@nokia.com
[(2)] Sorbonne Université, CNRS, LIP6, 4 place Jussieu, F-75005 Paris, France
[(3)] Université Paris-Saclay, IOGS, CNRS, Laboratoire Charles Fabry, 2 avenue Fresnel, F-91127 Palaiseau, France

**Abstract** *We designed a CV-QKD system with off-the-shelf components and established the feasibility of distributing 67.6 and 66.8 Mb/s secret key rates on average over a 9.5 km SMF link, using respectively probabilistically shaped 64 and 256 QAM, and relying on a novel analytical security proof.*

## Introduction

Continuous variable quantum key distribution (CV-QKD) based on Gaussian states is a promising solution to implement practical secure communications over insecure physical links, thanks to its compatibility with commercially available light sources and equipment[1],[2]. Experimental demonstrations of CV-QKD protocols have been performed and demonstrated the feasibility of distributing several Mb/s secret key rates[3],[4]. Several of those demonstrations employed discrete modulations, like QPSK[5], to allow for simpler practical implementation and post-processing. However, using low cardinality modulation formats like QPSK, results in significant reduction of the provable secret key rate since these constellations are not good approximations of a bidimensional Gaussian source. Recently we have demonstrated the feasibility of using polarization division multiplexed probabilistically shaped (PDM-PCS) 1024-QAM, with optimized Maxwell-Boltzmann distribution, and demonstrated the feasibility of average 38 Mb/s secret key rate distributed over 9.5 km of single-mode fiber (SMF). The PCS 1024-QAM is an accurate approximation of the Gaussian source; however, its generation requires fine-resolution digital-to-analog convertors. Since our first demonstration, a novel security proof has been proposed, which establishes the asymptotic security of discrete constellations with arbitrary cardinality[6]. Relying on this new proof, we demonstrate in the present work the experimental feasibility of using lower cardinality 64-QAM and 256-QAM formats, while increasing the symbol-rate compared to our previous work[7]. This allowed us to improve our previous secret key rate results to 67.6 Mb/s for PCS 64-QAM, and 66.8 Mb/s for PCS 256-QAM, over 9.5 km of SMF. Both rates are averaged over 200 blocks of $2.8 \times 10^6$ symbols, taking into account finite size effects.

## Continuous Variable QKD

In the CV-QKD protocol that we implement, the transmitter, Alice, prepares coherent states $|\alpha\rangle = |(p + iq)/2\rangle$, chosen at random from a discrete modulation. She sends them through an optical link to the receiver, Bob, who measures them using phase-diversity coherent detection. Afterwards, Alice and Bob reveal a fraction of their data on a public authenticated channel to evaluate the quantum channel parameters. These are used to infer the quantity of information obtained by Eve and compute the length of the final key. Then, Alice and Bob correct transmission errors through a step called reconciliation. Finally, they perform privacy amplification to extract the final secret key from their common data sets.

The quantum channel is characterized, after phase diversity coherent detection, by the equation[9]:

$$V_B = \frac{\eta T}{2} V_A + N_o + V_{el} + \xi_B$$

where $V_B$ is the variance of Bob's measured states in shot noise units (SNU), $V_A = 2\langle n \rangle$ is the variance of Alice's modulation, $\eta$ is the quantum efficiency of the receiver, $T$ is the channel transmittance, $N_o = 1$ is the shot noise variance, $V_{el}$ is the receiver's electronic noise variance and $\xi_B$ is the excess noise variance measured at Bob's site. The excess noise is attributed to a perturbation generated by Eve's attack. Since the best location for the attack would be at the output of Alice's lab, we are interested in the excess noise at the channel entrance $\xi_A = 2\xi_B/\eta T$.

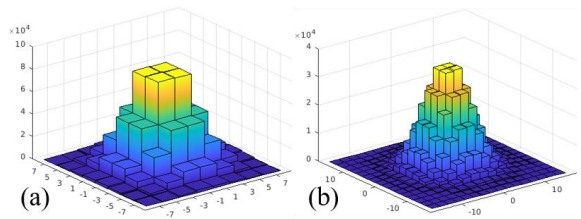

**Fig. 1** Histogram of **(a)** PCS 64-QAM with $\nu = 0.0749$, **(b)** PCS 256-QAM with $\nu = 0.0294$ for $10^6$ symbols.

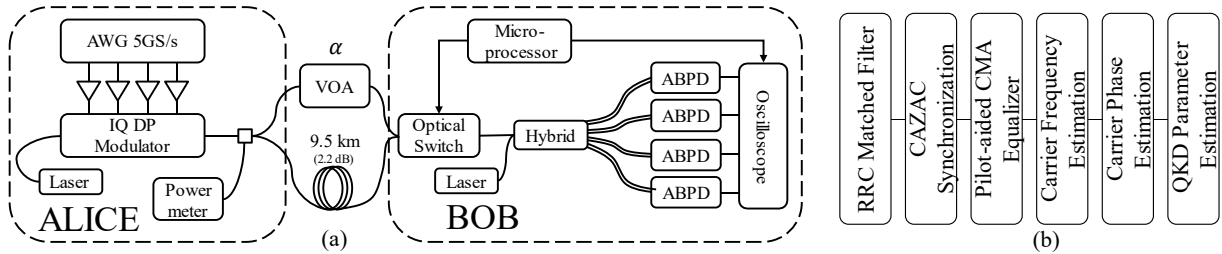

**Fig. 2 (a)** Experimental testbed featuring 10 kHz linewidth lasers sources, conventional I/Q dual polarization optical modulator, 5 GS/s arbitrary waveform generator, 180° hybrid, amplified balanced photodiodes (ABPD), 1GHz oscilloscope with 5 GS/s sampling rate. **(b)** Standard DSP procedure.

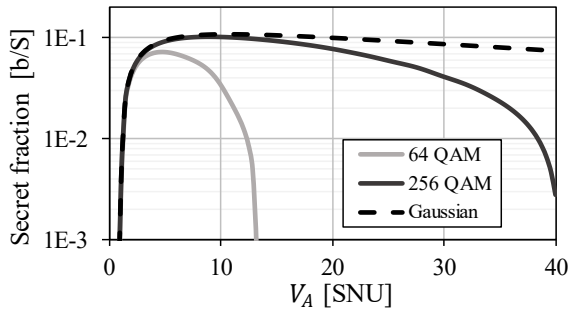

**Fig. 3** Secret fraction in bits/symbol vs $V_A$ where the PCS QAM probability distribution was optimized for each point, and with $T = -2.2$ dB, $\eta = 0.6$, $V_{el} = 0.1$ SNU, $\xi_B = 0.012$ SNU, $\beta = 0.95$.

**Modulation and security**

In this work the modulation of the coherent states is a probabilistic constellation shaping (PCS) QAM. A PCS QAM is a standard QAM with a Boltzmann-Maxwell probability distribution on the constellation points, with a free parameter $\nu$[8]. It is in fact a discrete modulation with a Gaussian-like probability mass function over the constellation points as illustrated in Fig. 1. For a given $V_A$, we choose $\nu$ such that it minimizes the trace distance between quantum states with PCS QAM and Gaussian modulation (cf. Fig. 2a in [7] and its accompanying text). In this way, we obtain the best approximation possible of the Gaussian modulation with the Maxwell-Boltzmann distribution over the square QAM lattice.

The security of protocols using PCS QAM can be established using explicit asymptotic lower bounds on the secret key rate for arbitrary modulations derived recently by Denys, Brown and Leverrier[6]. However, their approach uses a simplified model with ideal photodetectors. To take into account the quantum efficiency of the receiver $\eta$ as well as the electronic noise $V_{el}$, we combine their derivation of the covariance matrix of Alice's and Bob's states with a realistic model for the receiver[9]. Moreover, to tackle the issue of statistical errors in parameter estimation, we use a worst-case estimator for the excess noise[10], with $N = 2.8 \times 10^6$ symbols and security parameter $\epsilon = 10^{-8}$. Fig 3. gives the secret fraction in bits per symbol, computed for PCS 64 QAM and PCS 256 QAM, as well as the one for a Gaussian modulation. From this, we infer optimal values for $V_A$ around respectively 5 and 10 for 64 QAM and 256 QAM. By examining these theoretical curves, we observe that at its peak, the performance of the 256 QAM is very close to that of the Gaussian, with higher secret fraction than 64 QAM.

**Experimental system**

We assembled a CV-QKD experimental system with off-the-shelf telecom equipment, outlined in Fig. 2a. Alice uses a low-linewidth laser source tuned to 1550 nm. A 5 GSample/s arbitrary waveform generator (AWG) generates a dual polarization 600 MBaud signal with digital root-raised cosine (RRC) pulse shape with adjustable roll-off factor, which was optimized to 0.4[7]. To avoid low frequency noise from the hardware, both from Alice and Bob, the baseband signal was digitally upconverted by 500 MHz, so the useful bandwidth extends from 120 MHz to 880 MHz. Hence, the signal has no frequency component around DC. The four outputs of the AWG are fed to the analog part of a standard dual polarization I/Q optical modulator. The channel is a 9.5 km conventional SMF link with 2.2 dB characterized channel loss. Bob uses a 180° hybrid with four amplified balanced photodiodes to measure the received coherent states. The local oscillator (LO) at Bob's site has similar characteristics to Alice's laser source. The signal is sampled using a 1 GHz real-time oscilloscope with 5 GSample/s sampling rate. The sampled waveforms are stored for offline digital signal processing (DSP). Calibration of the shot noise is performed periodically using a fast optical switch to switch on and off the output of Alice.

**Digital Signal Processing**

The CV-QKD signal is interleaved in time with a deterministic sequence of QPSK pilot symbols, with 14 dB higher power than the QKD signal. The pilots are public information shared by Alice and Bob. They are used by the DSP to correct the optical channel impairments. Outlined in Fig. 2b is the DSP procedure, which uses only standard algorithms common in classical optical coherent transmissions[8],[11]. The waveforms sampled by the oscilloscope are assembled as two complex signals, one for each orthogonal polarization. A matched filter (RRC) is applied to the samples, and constant amplitude zero autocorrelation (CAZAC) sequences are used to synchronize the pilot sequence. Polarization demultiplexing is performed with a constant modulus algorithm (CMA) with coefficients updated on the pilots[11]. Carrier frequency estimation is performed using a periodogram method. Pilot-aided maximum-likelihood estimation applied on the pilots and combined with linear interpolation and optimization of filter length is used for carrier phase estimation.

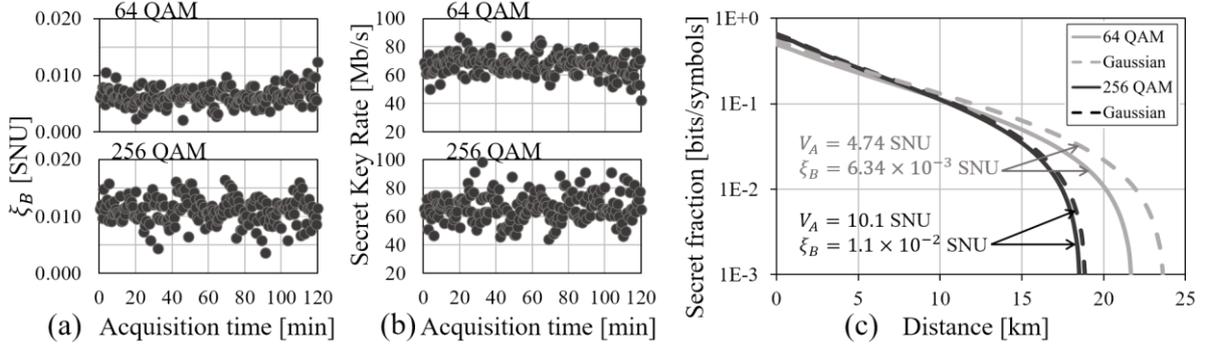

**Fig. 4. (a)** Excess noise variance at Bob's site $\xi_B$ vs. time over 9.5 km SMF (2.2 dB loss) for PCS 64-QAM and PCS 256 QAM, $N = 2.8 \times 10^6$ symbols per marker point and parameters of Tab. 1.
**(b)** Secret key rate vs. time for the same parameters, with finite size effects, and reconciliation efficiency $\beta = 0.95$.
**(c)** Secret fraction vs. distance, with parameters of Tab. 1, and $T = -2.2$ dB, $\eta = 0.6$, $V_{el} = 0.1$ and $\beta = 0.95$.

**Results**
Fig. 4a shows the measured excess noise variance $\xi_B$ at Bob's side in SNU for 200 blocks over a 2 hours long experiment, across 9.5 km SMF, for PCS 64-QAM and PCS 256-QAM. The excess noise variance is estimated using standard estimator of the variance for $N = 2.8 \times 10^6$ symbols. Fig. 4b shows the secret key rate for the corresponding excess noise values, computed with worst-case estimator, and a reconciliation efficiency $\beta = 0.95$. The formula for the secret key rate is

$$SKR = 2R_S(1 - R_{pilots})(\beta I_{AB} - \chi_{EB})$$

where 2 comes from the polarization multiplexing, $R_S$ is the symbol rate ($R_S = 600$ MBaud), $R_{pilots}$ the pilot rate ($R_{pilots} = 1/2$), $I_{AB}$ the mutual information between Alice and Bob and $\chi_{EB}$ is the Holevo information between Eve and Bob. Table 1 summarizes the measured averaged results.

**Tab. 1:** Measured averaged results

| Constellation | $\nu$ | $V_A$ | $\xi_B$ | SKR |
|---|---|---|---|---|
| / | / | SNU | SNU | Mb/s |
| PCS 64 QAM | 0.0749 | 4.74 | 6.34E-3 | 67.6 |
| PCS 256 QAM | 0.0294 | 10.1 | 1.10E-2 | 66.8 |

Assuming ideal hardware and DSP, 256 QAM results in higher SKR than 64 QAM at constant excess noise, as per Fig. 3. However, given the available commercial components and DSP suite that we use, the lower cardinality 64 QAM results in lower excess noise to the degree that the experimental SKR of 64 QAM is slightly better than 256 QAM. Moreover, as illustrated in Fig. 4, 64 QAM system can potentially reach longer distances, up to 22 km.

**Conclusion**
We investigated the use of PCS QAM modulations for CV-QKD using an analytical security proof for arbitrary modulations. We established the feasibility of block-averaged secret key rate of 66.8 Mb/s with PCS 256-QAM, and 67.6 Mb/s with PCS-64QAM over a 9.5 km SMF link, taking into account finite-size effects with worst case estimator and security parameter $\epsilon = 10^{-8}$. We believe it is possible to improve these results by increasing the block size and the overall stability of the experiment. This result contributes in improving the technology readiness level of the continuous-variable quantum key distribution systems, leveraging commercial components, standard DSP, and low-cardinality discrete constellations.

**Acknowledgement**
This work has received funding from the European Union's Horizon 2020 research and innovation under grant agreements No 820466 CiViQ and No 857156 OpenQKD. We thank Anthony Leverrier for useful discussions.

## References

[1] E. Diamanti and A. Leverrier, "Distributing secret keys with quantum continuous variables: principle, security and implementations," Entropy 17, 6072 (2015).

[2] H. Wang et. al, "High-speed Gaussian modulated continuous variable quantum key distribution with a local local oscillator based on pilot-tone-assisted phase compensation," *Opt. Express* **28**, 32882-32893 (2020).

[3] D. Huang, P. Huang, D. K. Lin, C. Wang, and G. H. Zeng, "High-speed continuous-variable quantum key distribution without sending a local oscillator," *Opt. Lett.* **40**(16), 3695–3698 (2015).

[4] F. Laudenbach, B. Schrenk, C. Pacher, M. Hentschel, C. F. Fung, F. Karinou, A. Poppe, M. Peev and H. Hübel, "Pilot-assisted intradyne reception for high-speed continuous-variable quantum key distribution with true local oscillator," *Quantum* 3, 193 (2019).

[5] T. A. Eriksson *et al.*, "Digital Self-Coherent Continuous Variable Quantum Key Distribution System," in *Optical Fiber Communication Conference (OFC) 2021,* paper T3D.5 (2020).

[6] A. Denys, P. Brown, and A. Leverrier, "Explicit asymptotic secret key rate of continuous-variable quantum key distribution with an arbitrary modulation," http://arxiv.org/abs/2103.13945.

[7] F. Roumestan, A. Ghazisaeidi, J. Renaudier, P. Brindel, E. Diamanti, P. Grangier, "Demonstration of probabilistic constellation shaping for continuous variable quantum key distribution," *Optical Fiber Communication Conference (OFC) 2021*, paper F4E.1 (2021).

[8] A. Ghazisaeidi et al., "Advanced C+L-Band Transoceanic Transmission Systems Based on Probabilistically Shaped PDM-64QAM," *J. Light. Technol.* **35**(7), 1291–1299 (2017).

[9] S. Fossier, E. Diamanti, T. Debuisschert, R. Tualle-Brouri, and P. Grangier, "Improvement of continuous-variable quantum key distribution systems by using optical preamplifiers," *J. Phys. B: At. Mol. Opt. Phys.* **42** (11), 114014 (2009).

[10] P. Jouguet, S. Kunz-Jacques, E. Diamanti, and A. Leverrier, "Analysis of Imperfections in Practical Continuous-Variable Quantum Key Distribution," *Phys. Rev. A* **86**(3), 032309 (2012).

[11] K. Kikuchi, "Fundamentals of Coherent Optical Fiber Communications," *J. Light. Technol.*, **34**(1), 157–179 (2016).